\documentclass[%
 reprint,
 amsmath,amssymb,
 prl,
]{revtex4-1}

\usepackage{graphicx}
\usepackage{dcolumn}
\usepackage{bm}
\usepackage{color}
\begin{document}

\title{Higher-order topological insulators in a crisscross antiferromagnetic model}
\author{Jinyu Zou}
\author{Zhuoran He}
\author{Gang Xu}
\email{gangxu@hust.edu.cn}
\affiliation{Wuhan National High Magnetic Field Center $\&$ School of Physics, Huazhong University of Science and Technology, Wuhan 430074, China}

\begin{abstract}
We present a $4'/m'$-respecting crisscross AFM model in 2D and 3D, both belonging to the $Z_2$ classification and exhibiting interesting magnetic high-order topological insulating (HOTI) phases. The topologically nontrivial phase in the 2D model is characterized by the fractional charge localized around the corners and the quantized charge quadrupole moment. Moreover, our 2D model also exhibits the quantized magnetic quadrupole moment, which is a unique feature compared with previous studies. The 3D system stacked from layers of the 2D model possesses the HOTI phase holding chiral 1D metallic states on the hinge, which corresponds to the Wannier center flow between the valence and conduction bands. The novel transport properties such as the half-quantum spin-flop pumping phenomena on the side surfaces of the HOTI phase is also discussed.
\end{abstract}

\maketitle

\textit{Introduction}.--Topological insulators (TIs) are distinctive quantum states characterized by the $Z_2$ invariant protected by time-reversal symmetry (TRS), which are stable under the continuous deformations of the band structure without closing the band gap \cite{Hasan2010,Qi2011}. A striking feature of the TI phase is the $(D-1)$-dimensional bulk-boundary correspondence, where the $D$-dimensional bulk is insulating but supports $(D-1)$-dimensional robust gapless boundary states \cite{Kane2005,Bernevig2006,Fu2007,ZhangHaijun2009}. In 2011, Fu \textit{et al.}~generalized the classification of topological materials to topological crystalline insulators (TCIs), where the gapless boundary modes are immune to local perturbations without breaking the crystalline symmetries \cite{Fu2011,Hsieh2012,Chiu2016,LiuChaoXing2015,Neupert2018,Song2018}. Recently, the $(D-1)$-dimensional bulk-boundary correspondence principle was generalized to a higher-order correspondence, and the topological phase is called higher-order topological insulator (HOTI) \cite{Benalcazar2017,BenalcazarPRB2017,Song2017,Langbehn2017,Schindler2018,SchindlerNP2018,WangArxiv2018}. In a $D$-dimensional $n$th-order TI ($n\leq D$), the $D$-dimensional bulk and the $D-1, \cdots, D-n+1$ dimensional boundaries are all gapped, but there are $(D-n)$-dimensional gapless boundary states protected by the crystalline symmetries. Especially, in the 2D second-order and 3D third-order TIs, the protected 0D corner states correspond to the quantized charge quadrupole and octupole moments \cite{Benalcazar2017}. Thus, these topologically nontrivial states turn out to be very rich in nature. To determine their topological numbers, a method has been established based on the band representations of the crystallographic space group at high-symmetry points in the Brillouin zone (BZ), which is a promising road to searching and constructing topological materials \cite{Bradlyn2017,Po2017,Zhang2019,Tang2019}.

According to the robust hinge states flowing between the valence and conduction bands, the 3D second-order TIs can be divided into helical HOTIs and chiral HOTIs \cite{Schindler2018}. The helical HOTIs preserve the TRS and support bidirectionally propagating gapless modes on the hinges. SnTe was the first predicted helical HOTI by first-principle calculations in Ref\cite{Schindler2018}, and the crystal bismuth was predicted and experimentally confirmed to possess the helical HOTI phase in Ref\cite{SchindlerNP2018}. The chiral HOTIs break the TRS and support unidirectionally propagating hinge states \cite{Miert2018,Kooi2018,Ezawa2018}. The Sm-doped Bi$_2$Se$_3$ \cite{Yue2019} and EuIn$_2$As$_2$ \cite{Xu2019} materials was proposed by first-principle calculations to exhibit the chiral HOTI phase, but the magnetic structure in Sm-doped Bi$_2$Se$_3$ is still under debate. Hence, the search for chiral HOTIs remains an important open question.

\begin{figure}
  \centering
  \includegraphics[width=8.5cm]{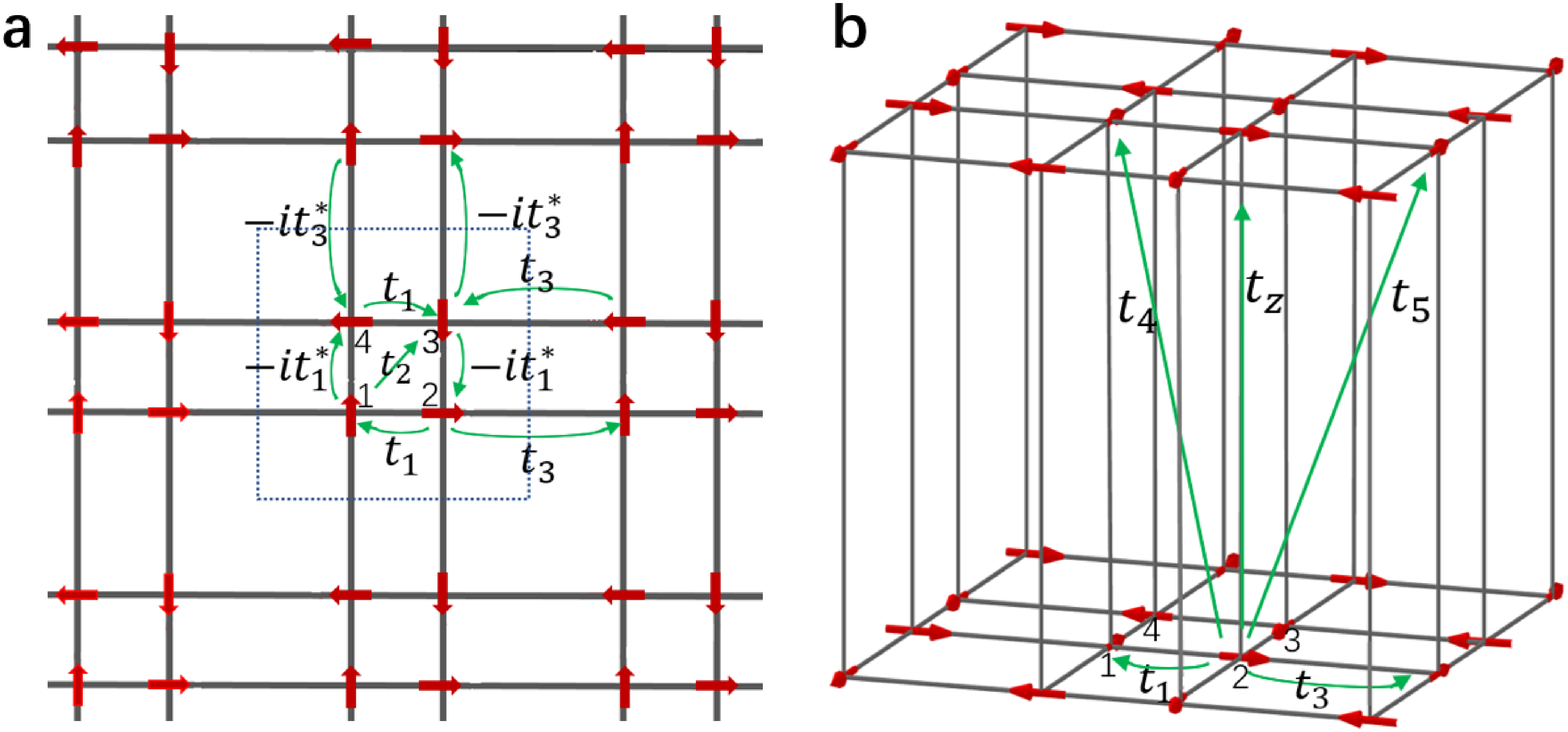}\\
  \caption{The configurations of the crisscross AFM model in 2D and 3D. (a) Schematic illustration of the 2D crisscross AFM model. The four sites in each unit cell are labeled by their corresponding numbers, and the pinned spin directions are represented by red arrows. The intra- and inter-cell hopping amplitudes are $t_1$ and $t_3$, respectively. (b) Schematic illustration of the 3D crisscross AFM model.}\label{models}
\end{figure}

From a theoretical point of view, chiral HOTIs can be constructed from the perspective of magnetic groups \cite{Okuma2019}, in which the TRS is broken but the combination of TRS with some crystalline symmetry is preserved. In this paper, we construct a crisscross antiferromagnetic (AFM) square lattice model satisfying $4'/m'$ magnetic point group (MPG), where the $m'=M_z T$ symmetry confines the spin polarizations in the $xy$-plane and then the $4'=C_{4z}T$ symmetry pins the spins in the directions as shown in Fig.~\ref{models}. Our 2D model can realize the nontrivial corner states with $1/4$ quantum magnetic quadrupole moment (MQM) and $1/2$ quantum charge quadrupole moment (CQM). By stacking the 2D lattice in $z$-direction, the system exhibits a novel 3D second-order TI phase, whose topological invariant can be determined from the band representations of $4'/m'$ at $C_{4z}T$-invariant points. The symmetry protected chiral states can exist robustly on the hinges of the 3D HOTI phase with insulating side surfaces, which can lead to topological magnetoelectric response and half-quantum spin-flop pumping behaviors.

\textit{2D Model}.--Here we introduce a crisscross AFM model on a 2D square lattice. As marked by the blue dotted square in Fig.~\ref{models}a, there are four sites in one unit cell. The spins on sites $1$ and $3$ point in the $\hat{y}$ and $-\hat{y}$ directions and the spins on sites $2$ and $4$ point in the $\hat{x}$ and $-\hat{x}$ directions. Obviously, such a 2D lattice model satisfies the MPG $4'/m'$ generated by $C_{4z}T$ and $PT$, where $T$ is the TRS and $P$ is the inversion symmetry. We set $c_{\alpha},c_{\alpha}^\dagger$ as the annihilation and creation operators on site $\alpha$ ($\alpha = 1,2,3,4$) in the spin directions given in Fig.~\ref{models}a, \textit{i.e} $\left|\uparrow_y\right> = (1,i)^T/\sqrt{2}$, $\left|\uparrow_x\right> = (1,1)^T/\sqrt{2}$, $\left|\downarrow_y\right> = (1,-i)^T/\sqrt{2}$ and $\left|\downarrow_x\right> = (1,-1)^T/\sqrt{2}$. Then $PT$ and $C_{4z}T$ act on these basis states as:
\begin{equation}\label{Generators}
\begin{split}
  &PT c^\dagger_{\alpha} (PT)^{-1} = (i)^{\alpha+2} c^\dagger_{\alpha +2}, \\
  &C_{4z}T c^\dagger_{\alpha} (C_{4z}T)^{-1} = e^{-i\frac{\pi}{4}}(i)^{\alpha+2} c^\dagger_{\alpha+1}.
\end{split}
\end{equation}
Yielding to the constrain of $4'/m'$ symmetries, the intracell nearest-neighbor hopping from site $2$ to site $1$ can be defined as $t_1 = \sqrt{2}\,_{\!}\lambda_{1\,}e^{-i\pi/4}$, with $\lambda_1$ being real, then the $C_{4z}T$ symmetry immediately requires the hopping amplitude from site $3$ to site $2$ to be $-it_1^*$. Such constraint also applies to the intercell nearest-neighbor hopping $t_3$, which can be written as $t_3 = \sqrt{2}\,_{\!}\lambda_{3\,}e^{-i\pi/4}$ with $\lambda_3$ being real. We note that the intracell next-nearest-neighbor hopping $t_2$ is zero due to the $PT$ symmetry.

\begin{figure}
  \centering
  \includegraphics[width=8.5cm, ]{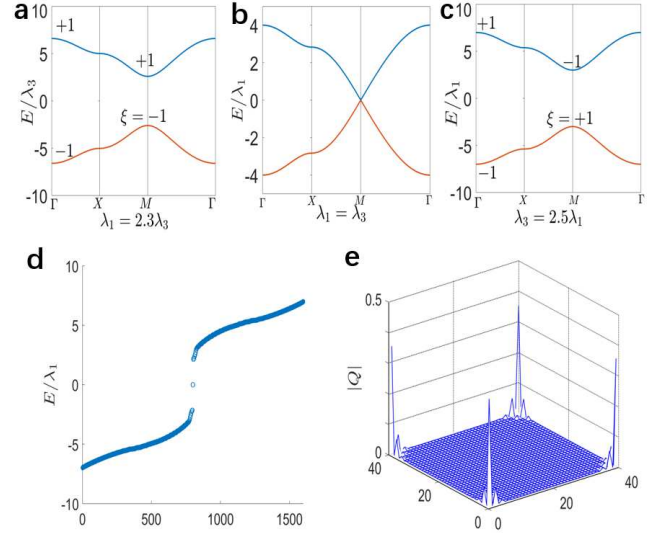}\\
  \caption{The band structure of the 2D crisscross AFM model and the corner states. (2a-2c) The band structure along high-symmetry lines with the parameters $\lambda_1 = 2.3\,_{\!}\lambda_3$ (trivial), $\lambda_1 = \lambda_3$ (gapless) and $\lambda_3 = 2.5\,_{\!}\lambda_1$ (nontrivial), respectively. Here $\xi=\pm 1$ labels the representation of the symmetry $S_4=\xi\,_{\!}e^{-i\frac{\pi}{4}\gamma_z}$ with the third Pauli matrix $\gamma_z$ at high-symmetry point $\Gamma$ and $M$. The band gap closes when $\lambda_1 = \lambda_3$ and band inversion happens when  $|\lambda_1| < |\lambda_3|$, leading to a nontrivial electronic structure. (2d) The energy levels for $20 \times 20$ unit cells of the nontrivial phase $\lambda_3 = 2.5\,_{\!}\lambda_1$. Four zero-energy modes emerge in the gap. (2e) The exponential distribution of the $e/2$ fractional charges carried by each corner state in $20 \times 20$ unit cells square.}\label{2D}
\end{figure}

By setting the lattice constants $|\mathbf{a}_1|=|\mathbf{a}_2|$ to be unity, the Hamiltonian under the constraints of the MPG $4'/m'$ in the momentum space with the basis $(c_{k1},c_{k2},c_{k3},c_{k4})^T$ are given by
\begin{equation}\label{HamiltonianAndEnergy2D}
\begin{split}
  H(k) &= (\lambda_1 + \lambda_3\cos{k_x})(\sigma_x+\sigma_y)\tau_z \\
       &+ (\lambda_1 + \lambda_3\cos{k_y})(\sigma_x+\sigma_y)\tau_y \\
       &- \lambda_3[\sin{k_x}(\sigma_x -\sigma_y)\tau_0 + \sin{k_y}(\sigma_x +\sigma_y)\tau_x],
\end{split}
\end{equation}
where each term is a direct product of Pauli matrices $\sigma_{x,y,z}$ and $\tau_{x,y,z}$, and $\tau_0$ is the $2\times 2$ identity matrix. Then the everywhere doubly degenerate dispersion relation is given by $E(k)= \pm 2\sqrt{\lambda_1^2 + \lambda_3^2 + \lambda_1\lambda_3(\cos{k_x} + \cos{k_y})}$, which separates the phase diagram of the system into four regions with two topologically distinct insulating phases: $|\lambda_1|>|\lambda_3|$ and $|\lambda_1|<|\lambda_3|$, and get a gapless phase transition state at $\lambda_1 = \pm \lambda_3$. We plot the band dispersion $E(k)$ along high-symmetry line for $\lambda_1 = 2.3\lambda_3$, $\lambda_1=\lambda_3$ and $\lambda_3 = 2.5\lambda_1$ in Figs.~\ref{2D}a--\ref{2D}c, respectively, and assume the Fermi level at zero. The two insulating phases shown in Fig.\ref{2D}a and Fig.\ref{2D}c can be distinguished by the position of the Wannier centers (WCs) of the occupied bands. In the trivial phase with $|\lambda_1| > |\lambda_3|$, the atomic orbitals on four sites hybridize to form $4$ Wannier orbitals and the WCs are located at the center of the unit cell. In the nontrivial phase with $|\lambda_1| < |\lambda_3|$, the WCs move to the corners of the unit cell, which means when cutting the infinite bulk into a square sample, zero-energy states will be left at the corners. Such a topological phase with insulating 2D bulk and 1D edge but 0D zero-energy modes is called 2D second-order TI \cite{Benalcazar2017}.

As discussed by Benalcazar et al \cite{Benalcazar2017,Liu2019}, the 2D second-order TI phase transition can also be understood from the change of the charge quadrupole moment (CQM), which is defined as
\begin{equation}\label{quadrupole}
  q_{xy} = \sum_{n=\mathrm{occupied}} P_x^n P_y^n /e,
\end{equation}
where $P_i^n = \frac{e}{2\pi}\int d^2k \langle u_n(k)|\partial_i|u_n(k)\rangle$ denotes the charge polarization of the $n$th band along the $i=x,y$ direction. With the unitary rotoinversion symmetry $S_4=PC_{4z}=(PT)(C_{4z}T)$, the polarization can be expressed by the representation of $S_4$ at the high-symmetry points $\Gamma$ and $M$ \cite{Schindler2018} as follows:
\begin{equation}\label{polarization}
  P_{x/y}^n = \frac{e}{2}\,_{\!}(\frac{\eta_M^n}{\eta_\Gamma^n}\ \mathrm{modulo}\ 2),
\end{equation}
where $\eta_{M/\Gamma}^n =\pm e^{\pm i\pi/4}$ denotes the $n^{th}$ band's eigenvalues of $S_4$ at $M/\Gamma$, and one get $P_x=P_y$ due to the $S_4$ symmetry. In Figs.~\ref{2D}a and \ref{2D}c, we have illustrated all bands' representation matrices of $S_4$ as $\xi\,_{\!}e^{-i\pi/4\gamma_z}$ with $\xi = \pm 1$. Explicitly, for the HOTI phase in Fig.\ref{2D}c, two occupied states' $S_4$ representation at $M$ and $\Gamma$  points are $\eta_{M} =\{ e^{i\pi/4}, e^{ -i\pi/4}\}$ and $\eta_{\Gamma} =\{ -e^{ i\pi/4}, -e^{ -i\pi/4}\}$ respectively. Thus one can obtain a nonzero CQM $q_{xy} = e/2$ in the HOTI phase $|\lambda_1| < |\lambda_3|$.

The nonzero CQM implies the fractional corner charges \cite{Benalcazar2017}, which corresponding to the localized corner sates. In Fig.~\ref{2D}d, we plot the energy levels of a square sample of $20\times 20$ unit cells in the nontrivial phase with $\lambda_3 = 2.5\,_{\!}\lambda_1$, where four zero-energy corner states related by $C_{4z}T$ all appear in the gap. At half filling ($2e$ per unit cell, $2L^2$ total electrons), the four corner states will share two electrons and each corner state carries a fractional charge $e/2$, which is exponentially distributed around the corner, as shown in Fig.\ref{2D}e.

Besides the CQM and localized fractional corner charges, our model also exhibits the magnetic quadrupole moment (MQM), which is a unique property compared with previously studied 2D HOTI models constructed from the $M_x$ and $M_y$ symmetries \cite{Benalcazar2017} or the $S_4$ rotoinversion group \cite{Miert2018,Ezawa2018}. In a magnetic lattice model, we can define the MQM in a unit cell as $\varrho_{ij} = \frac{1}{2}\,_{\!}(r_im_j+r_jm_i), i,j=x,y$, where $\mathbf{r} \ \mathrm{modulo}\  \mathbf{a}_1, \mathbf{a}_2$ is the position of the orbital carrying the magnetic moment $\mathbf{m}$. In our model, the magnetic moments are confined in the $xy$-plane by the $M_zT$ symmetry, and the $2$-fold rotation $C_{2z} = (C_{4z}T)^2$ requires that both the $x$ and $y$ coordinates of the WCs must be either $0$ or $1/2$ in units of the lattice constant. Thus, in our HOTI phase, when the WCs are not at $0$, the quantized MQM tensor is given by
\begin{equation}\label{MQM}
  \varrho_{ij} = \frac{1}{4}\,_{\!}g\,_{\!}\mu_B\left(
             \begin{array}{ccc}
               1 & -1 & 0 \\
               -1 & -1 & 0 \\
               0 & 0 & 0 \\
             \end{array}
           \right)\!,
\end{equation}
where $g$ is the electron Lande factor and $\mu_B$ is the Bohr magneton.

\begin{figure}
  \centering
  \includegraphics[width=9cm]{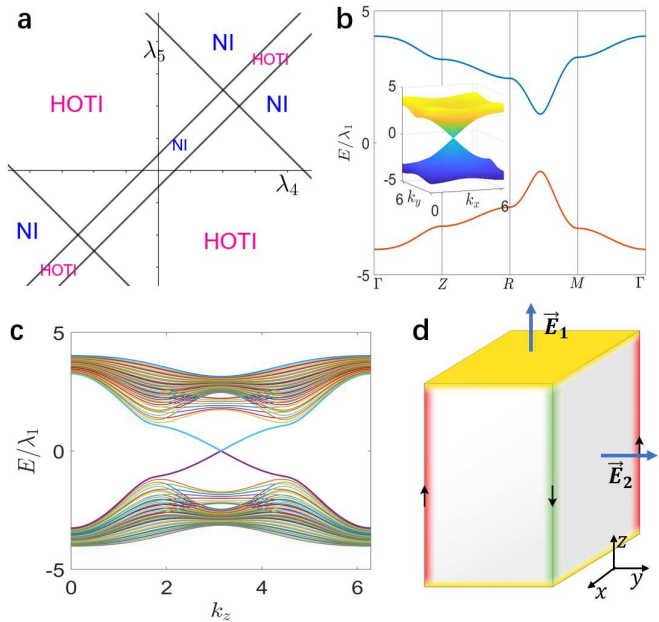}\\
  \caption{The band structure of the 3D crisscross AFM model, the Dirac surface states, and the chiral hinge states. (3a) Phase diagram in $\lambda_4/\lambda_1, \lambda_5/\lambda_1$ space with $\lambda_3 = 0.8\lambda_1$. (3b) The band structure along high-symmetry line for a HOTI with $\lambda_3 = 0.8\lambda_1, \lambda_4 = 0.41\lambda_1, \lambda_5 = -0.3\lambda_1, \lambda_z = 0.55\lambda_1$, where $\Gamma = (0,0,0), Z = (0,0,\pi), R = (\pi,\pi,\pi), M = (\pi,\pi,0)$. The inset shows the band structure of the HOTI with open boundary in $z$ direction and periodical boundary in $x,y$ direction. Two Dirac cone come from up and down surface degenerate at Fermi level because of the combined $PT$ symmetry. (3c) The band structure of the HOTI with open boundary in $x,y$ direction and periodical boundary in $z$ direction. Four edge states crossing the Fermi level correspond to the chiral hinge states. (3d) Schematic illustration of the Dirac surface states and chiral hinge states. The black arrows indicate the flow direction of the hinge states. $\vec{E}_1$ and $\vec{E}_2$ indicate the external electric field on $(001)$ and $(010)$ direction.}\label{3D}
\end{figure}

\textit{3D Model}.--As discussed above, the 2D systems for $|\lambda_1| > |\lambda_3|$ and $|\lambda_1| < |\lambda_3|$ correspond to distinct insulating phase, in which WCs locate at the center and corner of the unit cell, respectively. Hence it is nature to build a 3D tight-binding model where the $k_z=0$ and $k_z=\pi$ plane belong to different 2D phase. Such 3D model turns out to be a 3D second-order topological insulator with chiral hinge states, as shown in Fig.~\ref{3D}d. For this purpose, we construct a 3D lattice structure as illustrated in Fig.~\ref{models}b, where the interlayer hopping parameters are restricted by $4'/m'$ satisfying $t_4=\sqrt{2}\,_{\!}\lambda_{4\,}e^{-i\pi/4}$, $t_5=\sqrt{2}\,_{\!}\lambda_{5\,}e^{-i\pi/4}$, and $t_z=-i\lambda_z$, with $\lambda_4$, $\lambda_5$, and $\lambda_z$ being all real. Since the real part of $t_z$ only gives an overall shift of the energy bands, we take $t_z$ to be purely imaginary. Setting the interlayer distance as unity, the Hamiltonian in 3D momentum space takes the form as
\begin{equation}\label{HamiltonianAndEnergy3D}
\begin{split}
  H_\mathrm{3D}(k) &= (\lambda'_1 + \lambda'_3\cos{k_x})(\sigma_x+\sigma_y)\tau_z \\
       &+ (\lambda'_1 + \lambda'_3\cos{k_y})(\sigma_x+\sigma_y)\tau_y \\
       &- \lambda'_3[\sin{k_x}(\sigma_x -\sigma_y) \tau_0 + \sin{k_y}(\sigma_x +\sigma_y)\tau_x]\\
       &+ 2\lambda_z\sin{k_z}\,_{\!}\sigma_z\tau_0, \\
\end{split}
\end{equation}
where the effective parameters $\lambda'_1 (k_z)=  \lambda_1 + 2\lambda_4\cos{k_z}$ and $\lambda'_3 (k_z) =  \lambda_3 + 2\lambda_5\cos{k_z}$ depend on $k_z$.

Obviously, the $4'/m'$ MPG is preserved in the $k_z = 0$ and $\pi$ planes, in which the 3D Hamiltonian can hence be reduced to 2D model described by the formula (\ref{HamiltonianAndEnergy2D}) with parameters $\lambda'_1 = \lambda_1 \pm 2\lambda_4$ and $\lambda'_3 = \lambda_3 \pm 2\lambda_5$. Depending on whether the $k_z=0$ and $\pi$ planes holding the same or different 2D topological phases, our 3D model can exhibit either the normal insulator (NI) phase or the second-order TI phase, which can be distinguished by a topological number $v$ defined through the $S_4$ symmetry eigenvalues as
\begin{equation}\label{Z2}
  (-1)^v = \frac{\xi_R\xi_M}{\xi_Z\xi_\Gamma}.
\end{equation}
Here, $\{\xi_\beta e^{i\pi/4}, \xi_\beta e^{-i\pi/4}\}$ ($\xi_\beta = \pm1$ and $\beta = R,M,Z,\Gamma$) are the $S_4$ eigenvalues of the occupied bands at the high-symmetry points. The NI phase has $v=0$ and the HOTI phase has $v=1$. For the $v=1$ phase, symmetry protected chiral hinge states should exist robustly respecting to all the $C_{4z}T$ symmetry preserving perturbations. For example, one can add opposite 2D Chern insulators on the $x$-terminated and $y$-terminated surfaces of the HOTI, respectively. Such perturbation will change the number of edge states on each hinge by an even number. In that sense, the chiral HOTI phase here is classified by a $Z_2$ topological invariant, which is different from EuIn$_2$As$_2$ that belongs to $Z_4$ classification with $v=2$ standing for HOTI and $v=1,3$ standing for TI phase \cite{Xu2019}.

To drive the system to the HOTI phase, the parameters should satisfy
\begin{equation}
\begin{split}
[(\lambda'_1(0))^2-(\lambda'_3(0))^2][(\lambda'_1(\pi))^2-(\lambda'_3(\pi))^2] < 0.
\end{split}
\end{equation}
Thus the phase diagram is divided by the hyperplanes in the parameter space $(\lambda_1, \lambda_3, \lambda_4, \lambda_5)$. In Fig.~\ref{3D}a, we plot a section of the phase diagram as an example, by assuming $\lambda_1 = 1$ and $\lambda_3=0.8$. Evaluate the band dispersion $E(k) = \pm 2\sqrt{\lambda'^2_1 + \lambda'^2_3 + \lambda'_1\lambda'_3(\cos{k_x} + \cos{k_y}) + \lambda_z^2\sin^2k_z}$, the energy gap closes at one of the $S_4$ high-symmetry points ($\Gamma$, $M$, $R$, or $Z$), when the topological phase transition occurs.

The band dispersion $E(k)$ along high-symmetry line for the HOTI phase is plotted in Fig.~\ref{3D}b. As shown in Figs.~\ref{3D}c--\ref{3D}d, when the open boundary is set in $x$ and $y$ direction, the 2D surfaces are all gapped but the unidirectionally metallic states can survive on the surfaces intersecting hinges, which are protected by the $C_{4z}T$ or $S_4$ symmetry \cite{Schindler2018}. However, if the open boundary is set in $z$ direction, as shown in the inset of Fig.~\ref{3D}b, the $(001)$ surface band structure is not gapped but exhibit a Dirac cone, which is immune to the perturbations that preserve the $C_{4z}T$ symmetry. To destroy the Dirac cone on $(001)$ surface, one can apply a magnetic field along the $z$ direction to break the $C_{4z}T$ symmetry but remain the $S_4$ symmetry, which will result in the connected hinge states in the hexahedron sample \cite{Ezawa2018}.

\textit{Discussion}.--
Finally, we would like to discuss some unique transport properties on the surface of the 3D HOTI phase. The chiral second-order TI phase can be viewed as the result of introducing TRS-breaking but $4'/m'$-preserving interactions to a first-order TI. Hence, the Dirac cones on the side surfaces $(010)$, $(0\bar{1}0)$, $(100)$, and $(\bar{1}00)$ are all gapped by mass terms, resulting in the massive Dirac fermion behavior. Due to the $C_{4z}T$ symmetry, the mass terms on neighbouring side surfaces have opposite signs. As a result, the 1D metallic states are unavoidable on the domain walls, \textit{i.e} the surfaces intersecting hinges, and are robust to any $C_{4z}T$-preserving local perturbations. It is known that massive Dirac fermions can provide a half-quantum Hall conductance given by $\frac{e^2}{2h}\mathrm{sgn}(m)$ \cite{Bernevig2013}, where $m$ is the Dirac mass. Therefore, our chiral second-order TIs can also be viewed as axion insulators \cite{Wieder2018,Essin2009,Mong2010,Yue2019} with the side surface holding half-quantum Hall conductances, which can lead to novel responses to external electric fields. If the electric field is applied along the $y$-direction (see $\vec{E}_2$ in Fig.~\ref{3D}d), the $(100)$ and $(\bar{1}00)$ surfaces will obtain opposite half-quantum Hall currents, which are connected by the surface states on $(001)$ and $(00\bar{1})$ surfaces. Such transport phenomena is a natural result of the topological magnetoelectric effect in the axion insulator with the effective action
\begin{equation}\label{TME}
  S_\theta = \frac{\theta e^2}{4\pi^2}\int dt d^3\mathbf{r}\mathbf{E}\cdot \mathbf{B}
\end{equation}
where the axion angle $\theta=\pi$ \cite{Essin2009}. More interestingly, if the electric field is applied in the $z$-direction (see $\vec{E}_1$ in Fig.~\ref{3D}d), the charge will be pumped from two diagonal hinges to the other two hinges. Considering that the spin polarization directions of four hinge states are pinned in the $xy$-plane by $M_zT$ symmetry and must satisfy the $C_{4z}T$ symmetry, the spin direction will be deflected to the perpendicular direction through the pumping procedure on a side surface. Therefore, with the electric field applying in the $z$-direction, one can observe the half-quantum spin-flop pumping phenomena on the side surfaces.

The authors thank C.-X.Liu, X.Liu, Z.-X.Liu and Z.-D.Song for helpful discussions. The authors acknowledge the support by the Ministry of Science and Technology of China (2018YFA0307000), and the National Natural Science Foundation of China (11874022).
\bibliography{HOTI}
\end{document}